\documentclass[12pt, a4paper]{article}

\usepackage[version=3]{mhchem} 
\usepackage[T1]{fontenc}       
\usepackage{amsmath}
\usepackage{verbatim}
\usepackage{tabularx}
\usepackage[numbers,sort]{natbib}
\usepackage[english]{babel}
\usepackage{array}
\usepackage{graphicx}
\usepackage{comment}
\usepackage{color,soul}
\usepackage{authblk}

 \title{Stability of Surface Nanobubbles: A Molecular Dynamics Study}

\author[1]{Shantanu Maheshwari}
\author[1]{Martin van der Hoef}
\author[2,1]{Xuehua Zhang}
\author[1,3]{Detlef Lohse\thanks{d.lohse@utwente.nl}}
\affil[1]{Physics of Fluids, Mesa+ Institute, and J. M. Burgers Centre for Fluid Dynamics, University of Twente, P.O. Box 217, 7500 AE, Enschede, The Netherlands.}
\affil[2]{Soft Matter and Interfaces Group, School of Engineering, RMIT University, Melbourne, VIC 3001, Australia.}
\affil[3]{Max Planck Institute for Dynamics and Self-Organization, 37077, G\"{o}ttingen, Germany.}
\date{}

\begin{document}

\maketitle

 \begin{abstract}
The stability and growth or dissolution of a single surface nanobubble on a chemically patterned surface are studied
by Molecular Dynamics (MD) simulations of binary mixtures consisting of Lennard-Jones (LJ) particles.
Our simulations reveal how pinning of the three-phase contact line
on the surface can lead to the stability of the surface nanobubble, 
provided that the concentration of the dissolved gas is oversaturated.
We have performed equilibrium simulations of surface nanobubbles at different gas oversaturation levels $\zeta>0$.
The equilibrium contact angle $\theta_e$ is found to follow the
theoretical result of \citeauthor{lohse2015}\cite{lohse2015}, namely $\sin\theta_e = \zeta L/L_c$,
where L is the pinned length of the footprint and $L_c = 4\gamma/P_0$ a capillary length scale,
with $\gamma$ the surface tension and $P_0$ the ambient pressure.
For undersaturation $\zeta<0$ the surface nanobubble dissolves and the dissolution dynamics shows a
"stick-jump" behaviour of the three-phase contact line. 

\end{abstract}

\section{Introduction}

Surface nanobubbles are gaseous nanoscopic entities on an immersed surface that are less than 1 $\mu$m in height \cite{lohse2015rmp}. The presence of surface nanobubbles has significant implications to many physical and chemical processes at solid-liquid interfaces, for example, in fabrication of bubble-templated nanostructures \cite{fan2005am,hui2009,darwich2011}, propelling microscopic swimmers fuelled by catalytic reactions \cite{paxton2004,wilson2012}, light conversion by plasmonic effects\cite{fang2013}, heterogeneous cavitation under ultrasound\cite{belova2011,belova2013}, onset of boiling on microscopic scale \cite{zhang2014prl}, and flotation in mineral processing \cite{hampton2009me,hampton2010}. Hence the stability and the dynamics of surface nanobubbles is interesting from both a fundamental and from an applied point of view \cite{lohse2015rmp}. 

Experimental studies have shown that surface nanobubbles have long lifetime, even for only a mild oversaturation level supplied to the system\cite{parker1994,stevens2005,lou2000,zhang2007prl,zhang2013JSR}. When a bubble dissolves or grows,  the morphological features of the bubble suggest a stick-slip motion of the contact line \cite{zhang2013langmuir,german2014}. The origin of pinning is attributed to intrinsic features of the solid surface. Different theories were proposed to explain the stability of  surface nanobubbles \cite{ducker2009,brenner2008,weijs2013prl,petsev2013}. Recently 
\citeauthor{lohse2015}\cite{lohse2015} provided the exact calculation for the stability of a single surface nanobubble\cite{lohse2015}.  Their derivation assumes pinning and builds on, not more than only the diffusion equation,
Henry's law, and the Laplace equation. Their calculation reveals that contact line pinning  and the gas oversaturation $\zeta>0$ in the bulk liquid are crucial for the equilibrium of the single surface nanobubble to be stable. Here $\zeta= \frac{C_\infty}{C_s} -1$ is the gas oversaturation, $C_\infty$ the gas concentration, and $C_s$ the gas solubility. 
For a given gas oversaturation $\zeta>0$, there exists an equilibrium in which the outflux of gas molecules from the nanobubble due to the large Laplace pressure 
is compensated by the influx into the bubble due to gas oversaturation, i.e., there is no net flux. 
Though the flux of gas particles out or into the nanobubble is not spatially uniform along the interface as shown by
\citeauthor{weijs2012prl}\cite{weijs2012prl} and \citeauthor{yasui2015}\cite{yasui2015}, 
the {\it net} flux integrated over the whole interface is zero.
The contact angle of the surface nanobubble is not given by Young's equation, but determined by the equilibrium,

\begin{equation}
\sin\theta_e = \zeta\frac{L}{L_c}, \label{eq:eqangle}
\end{equation}
\nobreak where L is the pinned length of the footprint and $L_c = 4\gamma/P_0$ a capillary length scale,
with $\gamma$ the surface tension and $P_0$ the ambient pressure as shown in the sketch in Figure \ref{fig:simbox}.

Meanwhile, molecular dynamics (MD) simulations have provided important insight into the dynamics of surface nanobubbles, in particular under the conditions that are difficult to achieve in experiments. \citeauthor{weijs2012prl}\cite{weijs2012prl} performed  molecular dynamics (MD) simulations of surface nanobubbles without any heterogeneities and found that nanobubbles are not stable.
\citeauthor{liu2013jcp}\cite{liu2013jcp,liu2014jcp2} showed with the help of  kinetic lattice density functional theory and MD simulations 
that contact line pinning on geometrical heterogeneities also leads to stable surface nanobubble.  However, it remains unknown how exactly the oversaturation in the liquid and chemical patterns with a nanoscale dimension on a surface can mediate the stability of nanobubbles.
Moreover, given that it is extremely difficult to confirm eq \ref{eq:eqangle} experimentally, in this work we want to confirm
it with the help of MD simulations.

We will present our simulations of the dynamics (stability, growth, and dissolution) of a single surface nanobubble on a chemically patterned surface, in response to different oversaturation levels. The chemical patterns act as pinning sites enabling us to study the growth or dissolution of surface nanobubbles on a more realistic solid substrate,
exhibiting pinning and de-pinning of the contact line.

\section{Approach and methodology}
\begin{figure}
    \includegraphics[width=0.95\textwidth]{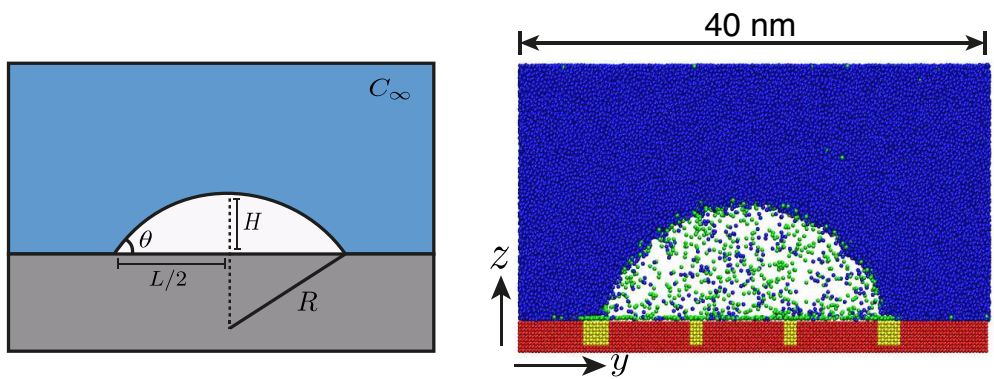}
  \caption{
  Left: Schematic of a single surface nanobubble on a hydrophobic surface and immersed in an oversaturated solution. L is the footprint diameter
  of the bubble, $\theta$ the contact angle, R the radius of the curvature of the bubble and $C_\infty>C_s$ the concentration of the gas far away from the interface. Right:
  A typical simulation box which consists of four kind of particles. Red particles ($\rm{S}$) form the solid surface, yellow particles 
  ($\rm{S}_P$) form pinning sites, blue particles (L), which are predominantly in the liquid phase, and green particles (G), which are
  predominantly in the gas phase. The nanobubble shown in the figure is cylindrical in shape 
  with the x-axis along the length of the cylinder.}
  \label{fig:simbox}
\end{figure}

Molecular Dynamics (MD) simulations were performed to simulate the nanobubble on a solid substrate 
for which we used the open source code GROMACS\cite{gromacs}. 
We have used four types of particles or molecules in our simulations: two types of solid particles ($\rm{S}$ and $\rm{S}_P$), which 
remain fixed in a fcc lattice during the whole simulation,
and two types of moving particles, Liquid (L) and Gas (G). The $\rm{S}_P$ particles form the pinning sites, and have 
different interaction strength towards the two types of moving particles. The L particles form a bulk liquid phase (and hence we refer to these
as "liquid particles") as the system temperature and pressure are below the critical point of L particles whereas the G particles form a bulk gaseous phase (to which we refer as "gas particles")
because the critical point for G particles is much below the thermodynamic conditions at which we are performing our simulations. 
A typical simulation box is shown in Figure \ref{fig:simbox}. 

The interaction between the particles is described by a Lennard-Jones potential:
\begin{equation}
\phi^{\rm{LJ}}_{ij} (r) = 4\epsilon_{ij}\Big[\Big({\frac{\sigma_{ij}}{r}}\Big)^{12} - \Big({\frac{\sigma_{ij}}{r}}\Big)^6\Big], \label{eq:lj}
\end{equation} 
\nolinebreak in which $\epsilon_{ij}$ is the interaction strength between particles $i$ and $j$, and $\sigma_{ij}$ is the characteristic 
size of particles.
The potential is truncated at a relatively large cut-off radius ($r_{c}$) of $5\sigma_{LL}$. The time step for updating the particle 
velocities and positions was set at  $dt = \sigma_{LL}\sqrt(m/\epsilon_{LL})/400$, where $m$ is mass of the liquid
particles and $\epsilon_{LL} $ is the Lennard-Jones interaction parameter for the liquid particles.
The time step has been chosen such that its value is sufficiently less than the shortest time scale available in the system\cite{frenkel2002}.
Periodic boundary conditions have been employed in all three directions, which suggest that same solid substrate is also present above the liquid layer.

Simulations have been performed in a $NPT$ ensemble where the temperature is fixed at $300 K$,
which is below the critical point for the Lennard-Jones parameters ($\sigma_{LL}$, $\epsilon_{LL}$) that we have set for the liquid particles.
Semi-isotropic pressure coupling is used for maintaining constant pressure which means that the simulation box can expand or contract
only in the z-direction to keep the pressure constant. This has been done to avoid the creation of gaps along the 
solid surface boundaries in the x and y directions.
Simulations were performed in a quasi-2D manner in which the length of the simulation 
box along the x-axis is considerably smaller than the lengths in the other two directions, which means the shape of the nanobubble is cylindrical
instead of spherical. Simulations of cylindrical nanobubbles save computation time; of course 
the effect of the modified shape on the 
Laplace pressure is taken into consideration.
The complete set of Lennard-Jones parameters that we have used in our
simulations are given in table \ref{tab:ljparam}. The typical system size is $5.6\times40\times24~\mbox{nm}^3$ in x, y and z direction respectively, where 
we note that the length of the z dimension changes during the simulation
to keep the pressure constant.

\begin{table}
\centering
\begin{tabular} { l | c c }

$\mathbf{i-j}$ & $\mathbf{\sigma_{ij}}$, \rm{nm} & $\mathbf{\epsilon_{ij}}$, \rm{kJ/mol} \\ \hline
$\rm{S}$-L & 0.34 & 1.8 \\
$\rm{S}_P$-L & 0.34 & 1.5 \\
$\rm{S}$-G & 0.40 & 2.0 \\
$\rm{S}_P$-G & 0.40 & 5.0 \\
L-G & 0.40 & 1.55 \\
G-G & 0.46 & 0.8 \\
L-L & 0.34 & 3.0 \\ \hline
\end{tabular}
\caption{Value of various LJ parameters used in the MD simulations} \label{tab:ljparam}
\end{table}

In the initial configuration, gas and liquid particles are arranged in a fcc lattice above the solid substrate. 
Initially, the liquid near the surface is highly oversaturated with gas particles in order to aid the bubble nucleation on the surface which decreases the equilibration time,
which is around $5\times10^{7}$ time steps (around $\sim$9 ns).
For simulations at different pressures, we have used the final configuration of the previous simulation as an initial configuration to save computation time.
In Figure \ref{fig:simbox} we show a typical equilibrium profile of a nanobubble on a chemical heterogenous surface.

\begin{figure}
  \includegraphics[width=0.99\textwidth]{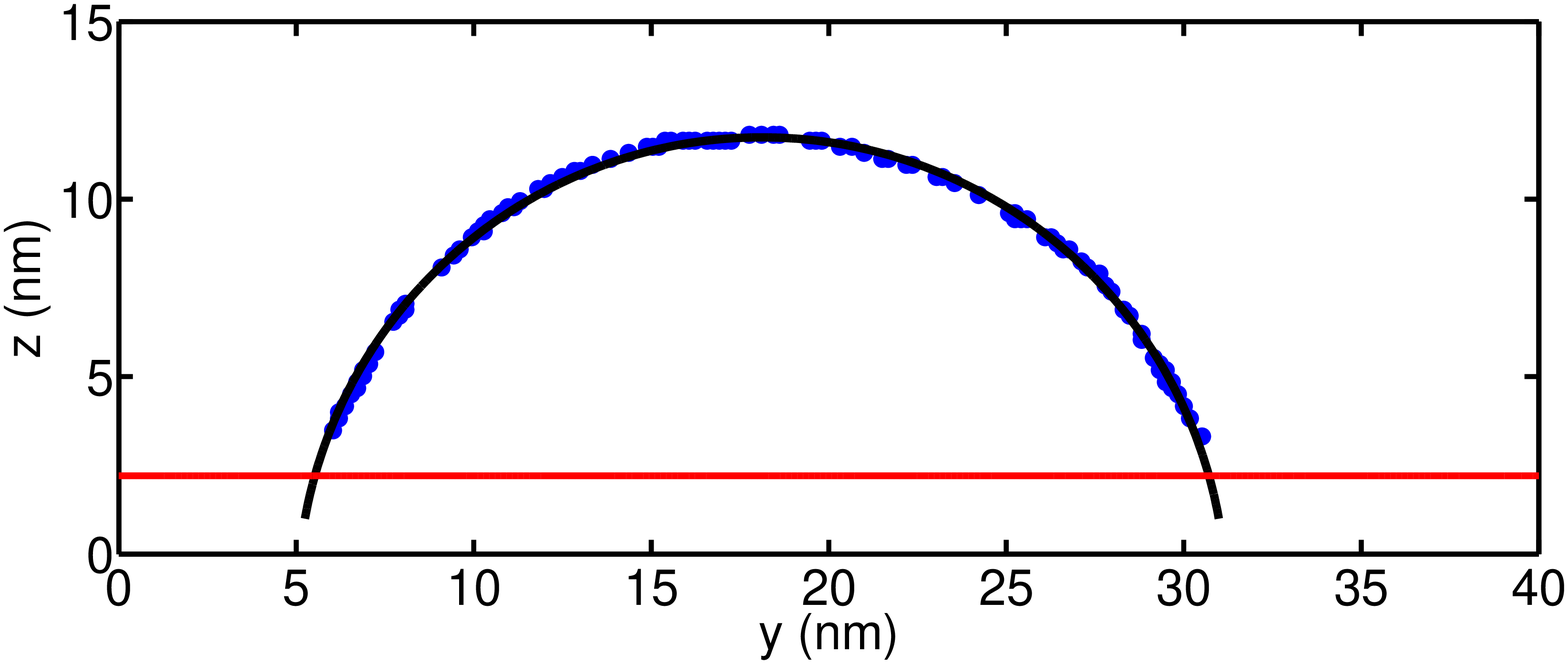}
  \caption{Example of a spherical cap fit (black line) to the iso-density contour of 0.5 (points) for a surface nanobubble}
  \label{fig:circlefit}
\end{figure}

After the equilibrium has been reached, the time-average density field of liquid particles is calculated,
correcting for the center of mass motion in the lateral direction.
Quantities like the radius of curvature of the bubble and the contact angle are obtained by
fitting a circle to the iso-density contour of $0.5$ of the normalised density field,
$\rho^{*}(r)$, defined as $\rho^{*}(r) = \frac{\rho(r)-\rho_{V}}{\rho_{L}-\rho_{V}}$, where $\rho_{V}$ and $\rho_{L}$ are the bulk vapour and liquid density,
respectively. Since the liquid very near to the solid 
substrate is subject to layering, we have excluded
the density field in the range of $2\sigma_{LL}$ from the substrate for the circular cap fitting.
From the intersection of the circular fit with the substrate, the contact angle and the radius of curvature are evaluated (see figure \ref{fig:circlefit}).
In order to study the time evolution for these quantities, we have calculated these time-averages for subsequent subsets of 25000 time steps.

The contact line is pinned by the pinning sites, which are formed by the solid particles with higher hydrophobicity ($\rm{S}_P$) as
shown by different colours in Figure \ref{fig:simbox}, which means that the interaction strength with gas particles is much higher than for the liquid particles.
Exact values for the interaction strengths are tabulated in table \ref{tab:ljparam}.
One can also notice from Figure \ref{fig:simbox} that the gas particles 
accumulate near the solid surface. This is due to the
hydrophobic nature of the solid surface which strongly attracts the gas particles, as discussed
in \cite{dammer2006}.

\begin{figure}
\centering
  \includegraphics[width=\textwidth]{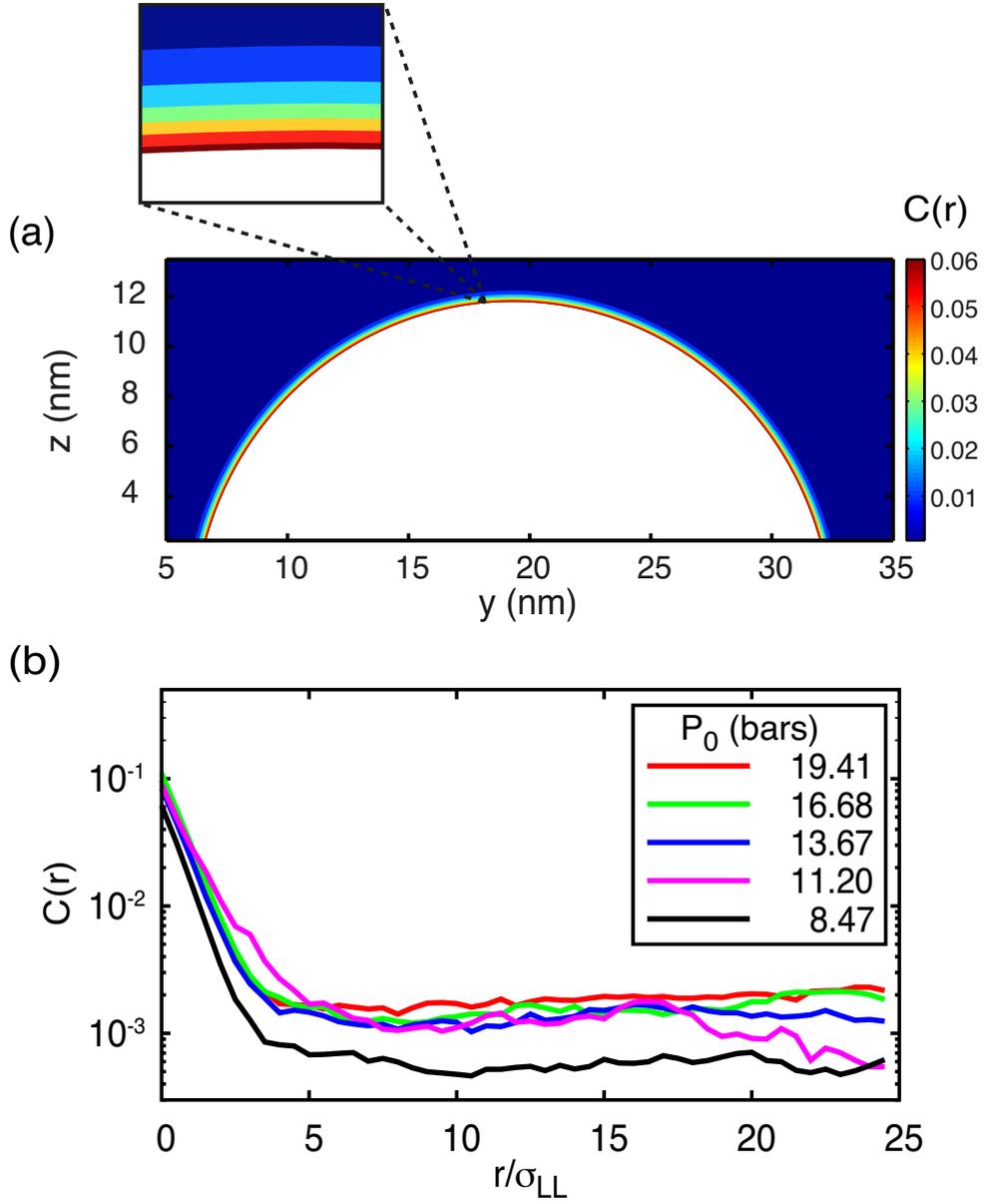}
  \caption{(a) Radial variation of concentration of gas particles around a nanobubble. The inset shows the concentration of gas particles near the interface.
  (b) Variation of concentration of gas against the distance from the interface at few pressures. The schematic in the inset shows the range of radial distance over which the concentration is plotted.
  }
  \label{fig:conc}
\end{figure}

\begin{figure}
\centering
  \includegraphics[width=0.99\textwidth]{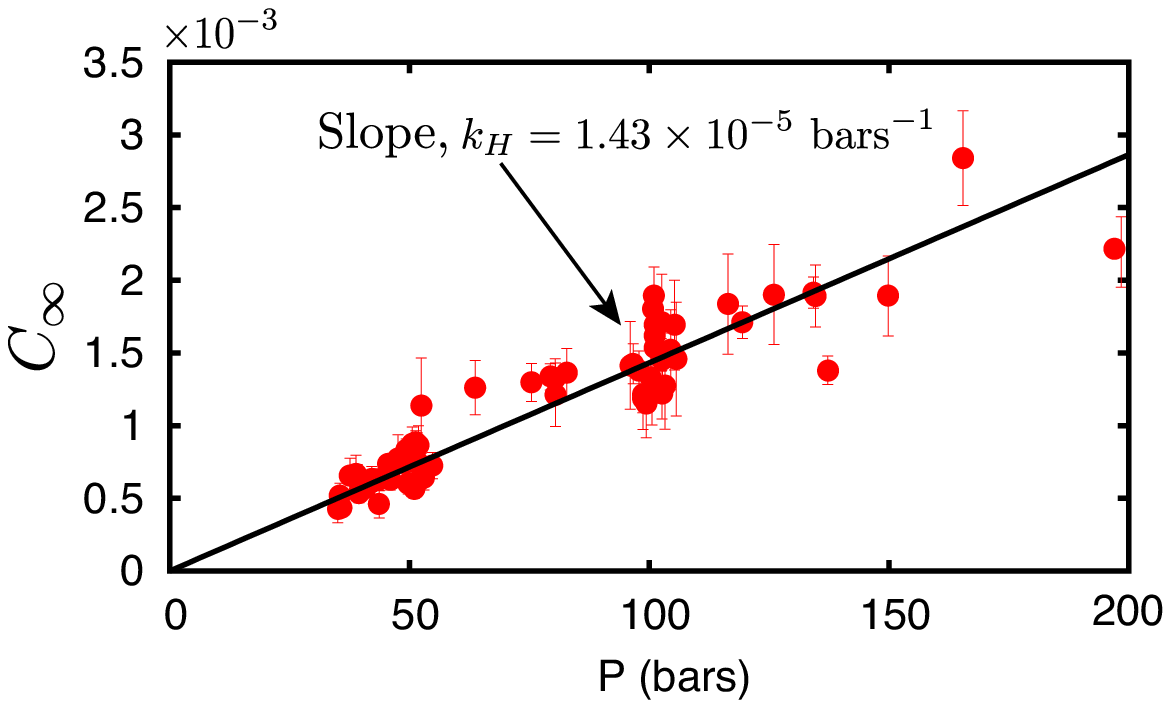}
  \caption{Concentration of gas particles in the bulk phase as measures in our simulations, for different gas pressures. The straight line suggests that the concentration of the gas particles indeed follows Henry's law
  $C_\infty = k_{H}P$.}
  \label{fig:henry}
\end{figure}

The oversaturation of gas particles in the bulk liquid is given by $\zeta= \frac{C_\infty}{C_s}-1$,  which involves the calculation of $C_\infty$ and $C_s$.
We have defined the concentration in this work as the ratio of gas particles to liquid particles in a certain amount of fixed volume, which is equivalent to the mole fraction of gas particles in the liquid.
The oversaturation $\zeta$ is controlled by changing the system pressure, since Henry's law dictates that the
solubility of gas in the liquid increases linearly with the system pressure. 
In our simulations, first a nanobubble is formed on the chemically heterogenous surface by a NVT simulation,
and then the pressure coupling is switched on, to keep the solubility of gas particles in the bulk liquid fixed. To change the solubility, and hence the oversaturation level,
the system pressure is slowly increased or decreased which respectively leads to the dissolution or growth of the bubble. For dissolution, the system pressure is increased which results in
the increase of solubility. Because of the finite system size, an increase in solubility leads to the migration of gas particles from the nanobubble to the bulk liquid, which leads to the dissolution
of the nanobubble. In order to determine $C_\infty$, we have divided the liquid domain into concentric shells of thickness $0.5\sigma_{LL}$, concentric around the
nanobubble, and then calculated $C(r)$, the ratio of
gas to liquid particles in each shell (see Figure \ref{fig:conc}). $C_\infty$ is then calculated by averaging 
$C(r)$ over the range of $7.5\sigma_{LL}$ to $25\sigma_{LL}$, where it is almost constant. In Figure \ref{fig:henry} we show $C_\infty$ as a 
function of the pressure of the gas phase, which at equilibrium is equal to the sum of the bulk pressure and the Laplace pressure. 
Our data are found to obey Henry's law, where we estimate Henry constant as $k_H = 1.43\pm0.027\times10^{-5}~\rm{bars}^{-1}$.
Note that the concentration of gas particles around the nanobubble is assumed to vary with radial distance only. In general, it varies in both radial and azimuthal direction
but it is a very weak function of the azimuthal angle when contact angle is around $\sim 90^{\circ}$\cite{popov2005}. In our simulations the contact angle is varying between $70^{\circ}$
and $90^{\circ}$, so the azimuthal angle dependence of the concentration field can be neglected.

\section{Results and Discussion}
\subsection{Pinning provides stability to surface nanobubbles}
\begin{figure*}
\center
  \includegraphics[width=\textwidth]{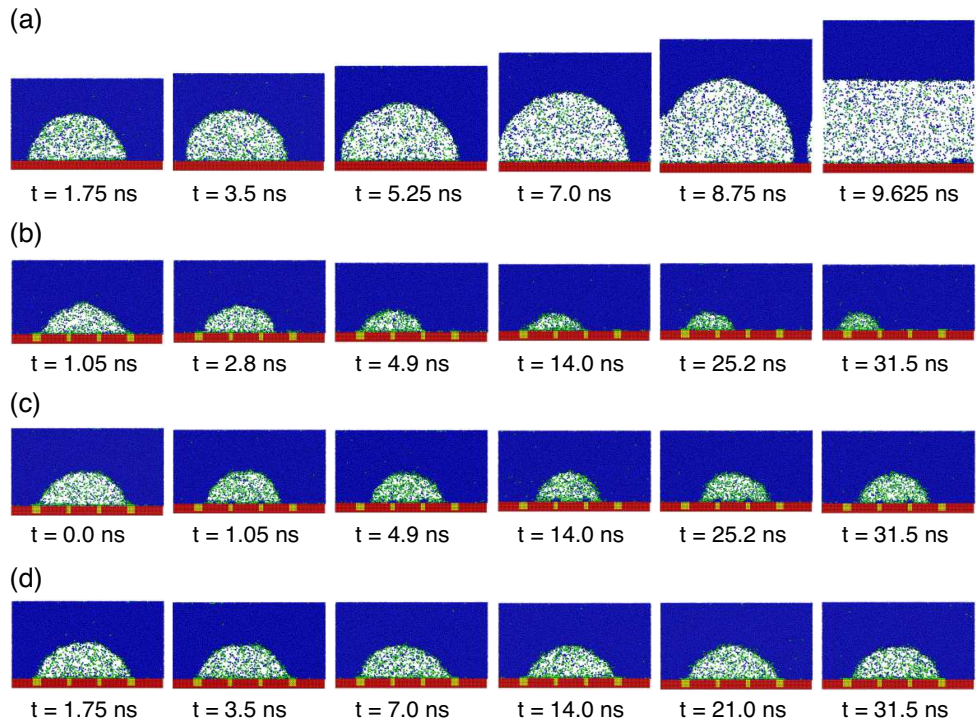}
  \caption{Time evolution of a surface nanobubble in a $NPT$ ensemble at a temperature of $300 K$ and pressure of 26 bars (a) without chemical heterogeneities and
  gas oversaturated liquid ($\zeta>0$), (b) with hydrophobic chemical heterogeneities and gas undersaturated liquid ($\zeta<0$), (c) with hydrophilic chemical heterogeneities and gas oversaturated liquid ($\zeta>0$), and
  (d) with hydrophobic chemical heterogeneities and gas oversaturated liquid ($\zeta>0$) surrounding the nanobubble.}
  \label{fig:stability}
\end{figure*}

\begin{figure*}
\center
  \includegraphics[width=0.8\textwidth]{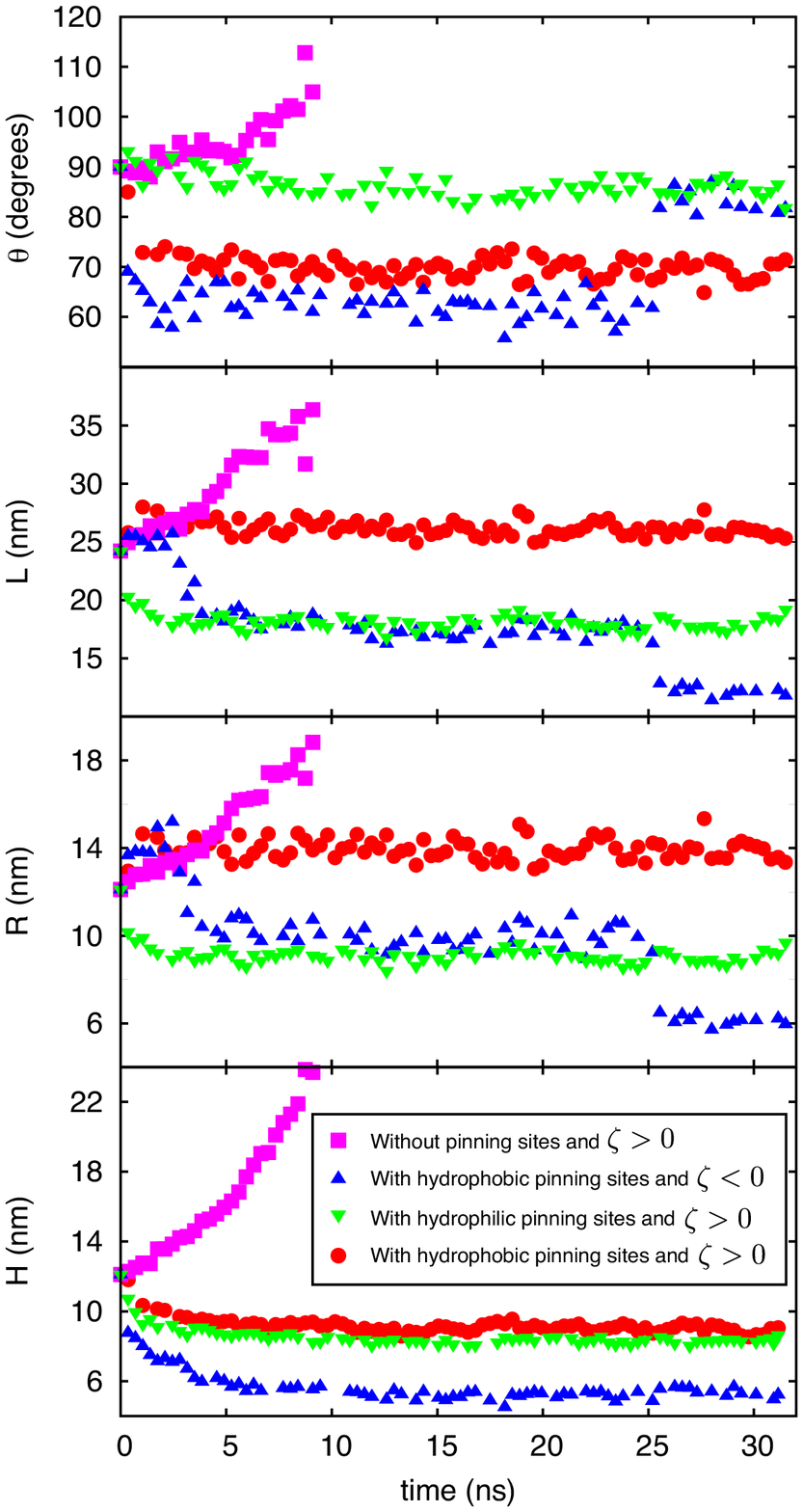}
  \caption{Time evolution of $\theta$, L, R and H of a surface nanobubble in the four cases shown in Figure \ref{fig:stability}.}
  \label{fig:timeevolution}
\end{figure*}

We start with testing the prerequisite conditions for the stability of nanobubbles given by \citeauthor{lohse2015}\cite{lohse2015}.
They theoretically showed that pinning of three-phase contact line by hydrophobic heterogeneities in a gas oversaturated liquid leads
to stability of nanobubbles. We now show systematically by performing MD simulations that these pinning conditions are indeed sufficient to 
get a stable surface nanobubble.
To this end we have performed four MD simulations at the same temperature and pressure, the time evolution of which is shown in Figures \ref{fig:stability}
(snapshots) and \ref{fig:timeevolution} ($\theta$, L , R, and H as function of time).
Figure \ref{fig:stability} (a) shows the time evolution of a nanobubble on a homogenous solid substrate, i.e. without any pinning sites, at a particular gas oversaturation, $\zeta>0$.
We find that the nanobubble in this simulation grows quickly due to large oversaturation of gas particles in the bulk liquid, so that after the few nanoseconds
it comes into contact with its periodic image, and ultimately forms a vapour liquid equilibrium with a planar interface.
It clearly shows the importance of heterogeneities on the solid surface without which it is impossible to form a stable nanobubble.
This behaviour is also shown in terms of the variation of $\theta$, L, R and H of the nanobubble with time in Figure \ref{fig:timeevolution}.
It is apparent from the plot that $\theta$, L, R and H increase rapidly with time and couldn't form a stable nanobubble.

Figure \ref{fig:stability} (b) shows the time evolution of a nanobubble on a chemical heterogeneous surface with hydrophobic pinning sites
at a particular gas undersaturation, $\zeta<0$. One can observe that the nanobubble slowly shrinks and dissolves with time. 
The nanobubble is not stable because the surrounding liquid is undersaturated with gas particles and it shows
the "stick-jump" mechanism during the dissolution, which means that the contact line remains pinned to the hydrophobic pinning sites.
This "stick-jump" dissolution mode is evident from Figure \ref{fig:timeevolution}, especially from the variation of L and R with time.
This kind of motion of the contact line in the dissolution of nanobubbles is very similar to the dissolution mode exhibited by microdrops dissolving in an
another liquid\cite{dietrich2015}.
This simulation clearly demonstrates the necessity of having gas oversaturation ($\zeta>0$) for the stability of nanobubbles.

Figure \ref{fig:stability} (c) shows the time evolution of a nanobubble on a chemical heterogeneous surface with hydrophilic pinning sites
at a particular gas oversaturation, $\zeta>0$. Hydrophilicity is introduced by interchanging the value of $\epsilon_{S_{P}G}$ and $\epsilon_{S_{P}L}$.
In this case we have not observed pinning of the three-phase
contact line and the nanobubble eventually dissolves, although the dissolution time is quite high.
Though it looks like from Figure \ref{fig:timeevolution} that a stable nanobubble is formed but in reality it slowly (compared to
the undersaturation case) dissolves.
Time scale for dissolution of a bubble for unpinned contact line case is given by the classical theory of \citeauthor{epstein1950}\cite{epstein1950,lohse2015,ljunggren1997}, namely
\begin{equation}
\tau_{life} \approx \frac{R_{0}^2\rho_g}{3DC_s}, \label{lifetime}
\end{equation}
\nobreak where $R_0$ is the radius of curvature of the bubble, $\rho_g$ the gas density, and $D$ the diffusion coefficient of gas particles in the liquid, which are extracted
from MD simulation of the nanobubble with hydrophilic pinning sites.
Accoding to this eq \ref{lifetime}, the lifetime of the nanobubble in our case is calculated as $\sim 2.9\mu s$. Simulating as long times as $\mu s$ is still very challenging for MD simulations; therefore we could not show 
the whole dissolution dynamics in this case.
However, one can infer from this simulation that only hydrophobic pinning sites lead to the pinning of the contact line and hence stable nanobubble.

Figure \ref{fig:stability} (d) shows the time evolution of a nanobubble on a chemical heterogeneous surface with hydrophobic pinning sites
at a particular gas oversaturation, $\zeta>0$. In this case, the nanobubble on the solid surface remains stable
throughout the simulation as chemical heterogeneities fixate the contact line, and the outflux due to Laplace pressure is balanced by the influx due to the
oversaturation of the gas particles in the bulk liquid. 
From Figure \ref{fig:timeevolution}, it is clear that in tis case $\theta$, L, R and H remain constant with time and form a stable equilibrium.
These four simulations confirm that hydrophobic chemical heterogeneities and gas oversaturation ($\zeta>0$) are necessary and sufficient conditions for forming a 
stable nanobubble. Note that also geometrical pinning sites lead to the stability of nanobubbles as already shown by \citeauthor{liu2014jcp2}\cite{liu2014jcp2}.

\subsection{Stable surface nanobubbles at equilibrium}
\begin{figure}
  \includegraphics[width=0.8\textwidth]{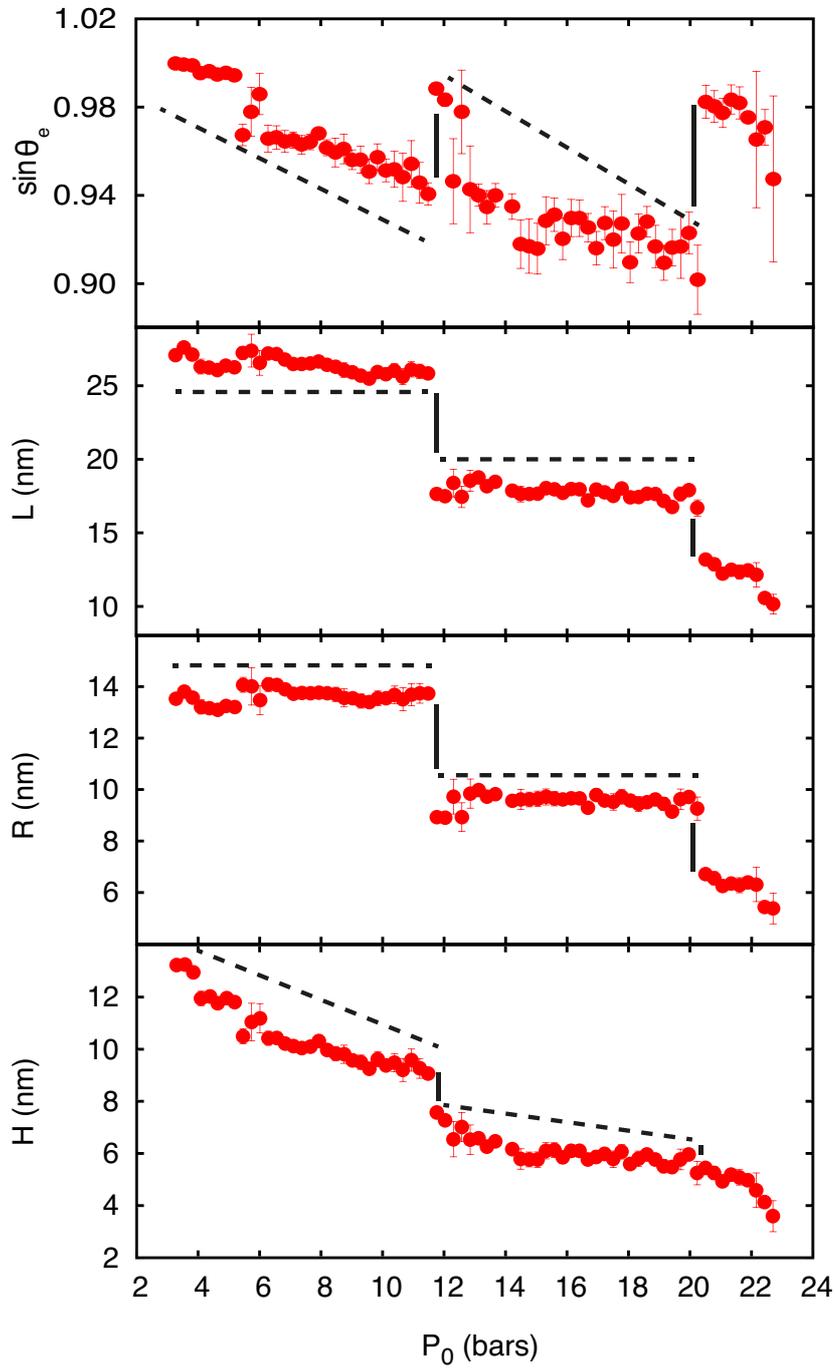}
  \caption{Variation of $\sin\theta$, footprint diameter (L) , radius of curvature of the nanobubble (R), and height of the nanobubble (H) with the bulk pressure $P_0$.
  Dashed lines show the sticking of the contact line between the two particular pinning sites
  and solid lines show the jump of the contact line from one pinning site to another. }
  \label{fig:jump}
\end{figure}

In this section we study the stable configurations of surface nanobubbles at equilibrium at various gas oversaturation levels.
The primary aim of this section is to validate the expression for the equilibrium contact angle of a surface nanobubble derived by \citeauthor{lohse2015}\cite{lohse2015}.
In the previous section we have qualitatively shown the necessary conditions for forming a stable nanobubble. Here, we systematically and quantitavely study the 
stable nanobubble at equilibrium with the help of MD simulations.
To this end, a nanobubble is nucleated on a chemical heterogenous surface with the help of local oversaturation 
at a particular temperature and pressure. After reaching equilibrium, the system pressure is slightly increased which results in an increase of the solubility of gas particles in the bulk liquid.
After increasing the pressure, the simulation was run for another 21 ns to ensure equilibrium of the system.
As explained before, the increased solubility (and the fixed number of the gas particles in the system) will lead to the dissolution of the surface nanobubble.  
This mechanism results in a sudden jump of the three phase contact line, at some particular pressure, towards a new pinning site, decreasing the
lateral length and suddenly increasing the contact angle of the nanobubble.
Figure \ref{fig:jump} shows the contact angle $\theta$, the footprint diameter L, the radius of curvature R of nanobubble, and its height H, as function of the bulk pressure $P_0$.
One can clearly observe the discontinuities in the variation which corresponds to 
the jump of the three-phase contact line over the pinning sites. 
The magnitude of change in the footprint diameter at discontinuities is equal to the distance between the pinning sites which confirms the "jump" of the
contact line on the pinning sites. 
We would like to emphasise that for each data point shown in Figure \ref{fig:jump}, the system is at equilibrium for that particular pressure. 

Note that $P_0$ is the bulk liquid pressure which is calculated in a liquid slab away from the interface. Although all the simulations are performed in a NPT ensemble,
where the pressure is controlled by semi-isotropic coupling, there is a large difference between the pressure set in GROMACS, and the bulk pressure $P_0$,
on which we elaborate in the appendix.

\begin{figure}
  \includegraphics[width=\textwidth]{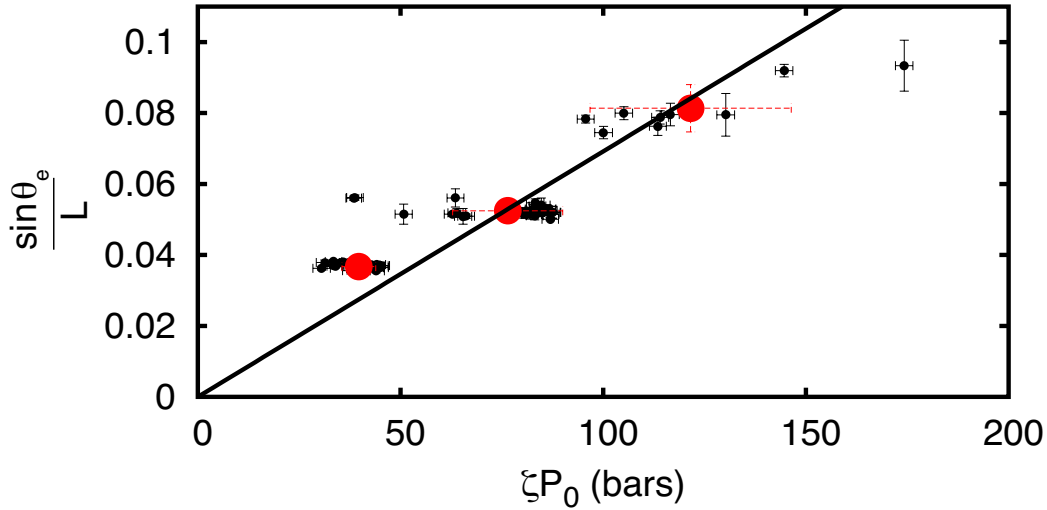}
  \caption{Variation of $\sin\theta/L$ against oversaturation the $\zeta{P_{0}}$. Small black data points represent the complete dataset shown in Figure \ref{fig:jump} while the three red points
    corresponds to the average of all data points for a nanobubble pinned on two particular pinning sites. The three data points refer to the three levels of L seen in Figure \ref{fig:jump}.}
  \label{fig:final_av}
\end{figure}

Theoretically, the equilibrium contact angle of a single surface nanobubble is given by eq \ref{eq:eqangle} $\sin\theta_e = \zeta L/L_c$. 
Note that here we had to modify the definition of $L_c$ from  $4\gamma/P_0$ to $2\gamma/P_0$, owing to the fact that 
for cylindrical bubbles the Laplace pressure is equal to $\gamma/R$, instead of the usual value $2\gamma/R$ that holds for spherical bubbles.
Since Henry's law is obeyed in our simulations, we can write $C_\infty = k_{H}P$ and $C_s = k_{H}P_0$, where
$P$ is the pressure inside the nanobubble and $P_0$ is the ambient bulk pressure. Note that now $P_0$ is calculated from the actual MD simulations and
not evaluated from the expression as given in the appendix.
We have plotted $\sin\theta_{e}/L$ against the oversaturation $\zeta{P_{0}}$ in Figure \ref{fig:final_av},
which shows that the equilibrium contact angle obeys eq \ref{eq:eqangle}. Note that there are some data points
which are slightly away from the theoretical line, which is mainly due to the large errors in the pressure calculation. 
Because of the relatively small system size, the instantaneous value for the pressure can deviate significantly from the mean which leads to large error bars\cite{gromacs}. 
Another reason for this deviation is the assumption of the surface tension being independent of pressure. 
The pressure inside the bubble directly affects the concentration of gas particles at the interface, which can have an influence on the magnitude of the surface tension.
The three points in the Figure \ref{fig:final_av} corresponds to the three radii of curvature or 
footprint diameter levels as shown in Figure \ref{fig:jump}. Each of the three points is the average of all the small data points for a nanobubble pinned on two particular pinning sites, which is also
shown in Figure \ref{fig:final_av}. 

Note that $\sin\theta_{e}/L$ is plotted against $\zeta{P_{0}}$ instead of $\zeta/L_{c}$, as we do not a priori know the value of the surface tension $\gamma$.
In fact we can use eq \ref{eq:eqangle}, to obtain an estimate for the surface tension $\gamma$. since the slope of the straight line in Figure \ref{fig:final_av} is $1/(2\gamma)$, which yields a surface tension $\gamma = 0.072\pm0.0048$ N/m.

\section{Conclusions}
Molecular dynamics simulations were performed for single surface nanobubbles on a chemical heterogenous surface and in a gas oversaturated liquid. Heterogeneities act as pinning sites
which fix the three-phase contact line and stabilise the equilibrium, in which the gas influx due to the gas oversaturation is balanced by the outflux due to the Laplace pressure. 
The theoretically predicted conditions\cite{lohse2015} for stable single surface nanobubble were tested with MD simulations, finding full confirmation. 
We have found that only hydrophobic heterogeneities lead to the pinning of the contact line and not the hydrophilic ones.
In addition, we have also studied the dissolution of surface nanobubbles on a chemical heterogenous surface in a gas undersaturated liquid. 
The dissolution of a surface nanobubble shows "stick-jump" behaviour which is very similar to the behaviour shown by the dissolution of microdrops in another liquid\cite{dietrich2015,dietrich2016,liu2013pre}.
We have also simulated the stable nanobubble at various levels of oversaturation and showed the variation of the contact angle, the footprint diameter, the radius of curvature and the height of the nanobubble 
with system pressure.
The equilibrium contact angle is following the analytical expression calculated in ref. \cite{lohse2015},
which herewith we thus confirm.
The next step will be to simulate interacting surface nanobubbles.

We thank Shantanu Choudhary for assistance with parallelizing some of the codes, Michiel van Limbeek for fruitful discussions and Sander Wildeman, Varghese Mathai, Poorvi Shukla for constructive comments on the manuscript,
SURFsara (funded by NWO) for providing the computational facilities for the simulations, and FOM
for financial support.

\section{Appendix}
In this section we describe the calculation of $P_0$, that we have used in Figures \ref{fig:jump}. We have performed NPT
simulations in which the pressure is prescribed. Yet, there is a clear difference between the input pressure and the actual bulk liquid pressure, $P_0$ that we have used.
The pressure in any MD simulation with pair interaction $\phi_{ij}(r)$ can be calculated from:

\begin{equation}
P = \rho{k_{B}}T + \frac{1}{3V}\left\langle\sum_{i<j}\frac{d\phi_{ij}}{dr_{ij}}.{r_{ij}} \right\rangle
\label{eq:press}
\end{equation}

\begin{figure}
  \includegraphics[width=\textwidth]{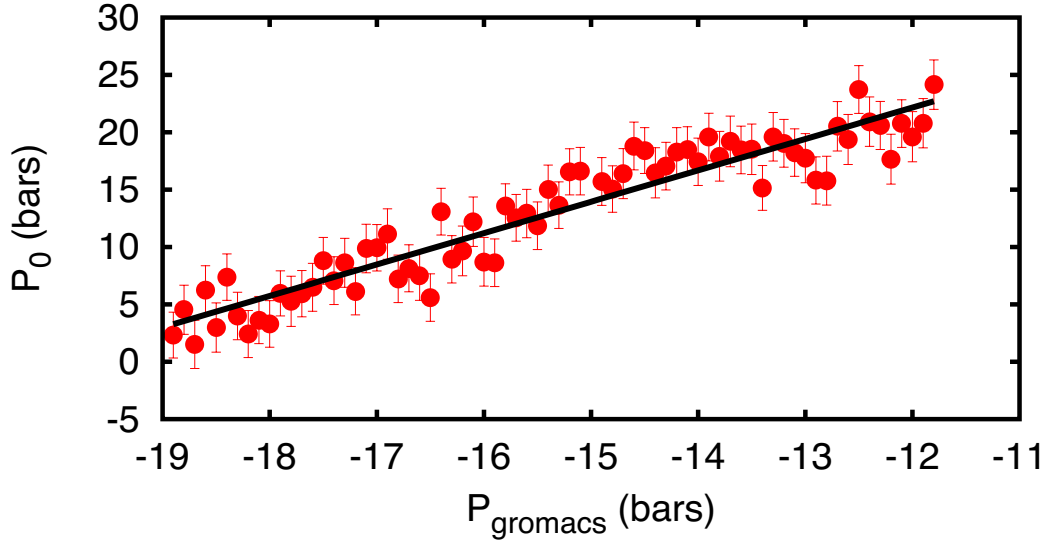}
  \caption{Variation of $P_0$ with input pressure in GROMACS}
  \label{fig:p0}
\end{figure}

As one can notice from the eq \ref{eq:press}, that calculation of pressure involves the interaction between all the possible pairs of particles which include
the interaction between solid and moving particles also, which is thus included in the pressure that is set for GROMACS.
However, due to the finite size of the system, this interaction between fixed solid and moving particles can have a huge effect on the overall pressure. 
In order to evaluate the "true" bulk pressure $P_0$, we have excluded this interaction from solid
particles, that is, $P_0$ is calculated by considering only the interactions in a rectangular slab
above the bubble. In the calculation of $P_0$, we have also considered the interaction between the pairs of $r_{ij}$ which are 
intersecting the boundary planes. Relation between the input pressure in GROMACS and $P_0$ is shown in Figure \ref{fig:p0}.

It can be noticed that $P_0$ is directly proportional to the input pressure in GROMACS and a straight line is fitted to get the relation between the two,
which is given as $P_0 = 2.73565{P_{gro}} + 54.9801$. $P_0$ in Figure \ref{fig:jump} is calculated from the input pressure using this relation, whereas
in Figure \ref{fig:final_av}, $P_0$ is the actual calculated value from the MD simulation. Now the question arises,
why we chose to use this relation instead of using $P_0$ directly calculated from the simulations. From Figure \ref{fig:p0}, it is clear that on an average $P_0$ is increasing 
linearly with input pressure but there are still some local fluctuations in the pressure which overcasts the jump in $\theta$, L, R, and H of surface nanobubble shown in Figure \ref{fig:jump}.

$P$, used in Figure \ref{fig:final_av} for calculation of $\zeta$, is the pressure inside the nanobubble which is calculated by considering 
the virial function of all pairs of particles within the bubble and half the virial function of the pairs whose 
$r_{ij}$ vector is intersecting the bubble boundary.

\bibliographystyle{unsrtnat}
\bibliography{references}

\end{document}